\documentclass[12pt,a4paper]{article}
\pdfoutput=1
\usepackage{indentfirst}
\usepackage{graphicx}
\usepackage{subfigure}
\usepackage{threeparttable}
\usepackage{array}
\usepackage{multirow}
\usepackage{amsmath}
\usepackage{amssymb}
\usepackage{hyperref}
\hypersetup{pdfauthor={Ruifeng Yuan, Sha Liu, Chengwen Zhong},pdftitle={A multiscale discrete velocity method for model kinetic equations}}

\begin{document}


\title{A multiscale discrete velocity method for model kinetic equations}
\renewcommand{\thefootnote}{\fnsymbol{footnote}}
\author{Ruifeng Yuan\footnotemark[1], Sha Liu\footnotemark[1], Chengwen Zhong\footnotemark[1]}
\footnotetext[1]{National Key Laboratory of Science and Technology on Aerodynamic Design and Research, Northwestern Polytechnical University, Xi'an, Shaanxi 710072, China}
\footnotetext{\emph{Email addresses:} xyrfx@mail.nwpu.edu.cn (Ruifeng Yuan), shaliu@mail.nwpu.edu.cn (Sha Liu), zhongcw@nwpu.edu.cn (Chengwen Zhong)}
\date{Dec 17, 2019}
\maketitle


\rule[-5pt]{\textwidth}{0.5pt}
\begin{abstract}
In this paper, authors focus effort on improving the conventional discrete velocity method (DVM) into a multiscale scheme in finite volume framework for gas flow in all flow regimes. Unlike the typical multiscale kinetic methods unified gas-kinetic scheme (UGKS) and discrete unified gas-kinetic scheme (DUGKS), which concentrate on the evolution of the distribution function at the cell interface, in the present scheme the flux for macroscopic variables is split into the equilibrium part and the nonequilibrium part, and the nonequilibrium flux is calculated by integrating the discrete distribution function at the cell center, which overcomes the excess numerical dissipation of the conventional DVM in the continuum flow regime. Afterwards, the macroscopic variables are finally updated by simply integrating the discrete distribution function at the cell center, or by a blend of the increments based on the macroscopic and the microscopic systems, and the multiscale property is achieved. Several test cases, involving unsteady, steady, high speed, low speed gas flows in all flow regimes, have been performed, demonstrating the good performance of the multiscale DVM from free molecule to continuum Navier-Stokes solutions and the multiscale property of the scheme is proved.
~\\

\end{abstract}
\rule[5pt]{\textwidth}{0.5pt}


\section{Introduction}
Rarefied gas flow simulation is an important field of computational fluid dynamics (CFD). Recent years, with the development of aerospace technology and micro-electromechanical system (MEMS), the rarefied flow simulation has attracted more and more attention. The discrete velocity method (DVM) \cite{Goldstein1989Investigations,Yang1995Rarefied,Mieussens2000Discretev,li2004Study,Titarev2007Conservative} , or also known as the discrete ordinate method (DOM), is a classical numerical method playing an important role and widely used in the area of rarefied flow simulation. Unlike another important method for rarefied flow simulation called direct simulation Monte Carlo (DSMC) \cite{Bird1994Molecular} which is based on the particle statistics, DVM is a deterministic method solving the Boltzmann equation or its model equations. In the conventional DVM, the operator splitting is used, the transportation term and the collision term of the governing equation are handled separately, which endows the scheme with a series of advantages like conciseness, easiness in programming, high efficiency and good accuracy in the simulation of high Knudsen (Kn) number. These assets just ensure the position of the conventional DVM in the area of rarefied flow simulation. However, on the other hand, due to the decoupling of the transportation term and the collision term, during the transportation process of the conventional DVM, particles are always transferring freely without any collision, causing an overly nonequilibrium particle velocity distribution when the numerical scale (i.e. the mesh size and the time step) is much larger than the kinetic scale (i.e. the mean free path and the mean collision time). This means that in the simulation of low Kn number where intensive particle collision occurs, the conventional DVM needs to adopt very small mesh size and time step, which requires huge amount of computation, otherwise the free-transport mechanism of the decoupled transportation term will make the result very dissipating. Thus, the conventional DVM can be hardly used to do fine simulations of flow in continuum regime.

The limitation of the conventional DVM makes it mainly applicable to single scale simulations where the numerical scale is comparable with the kinetic scale. However, nowadays in the field of CFD there are more and more important problems involving the continuum flow and the rarefied flow simultaneously in one flow field. These problems include but are not limited to the flow around the hypersonic vehicle which needs to travel between ground and near space atmosphere, or the flow in the reaction control system (RCS) of the spacecraft. The single-scale conventional DVM is not proper to deal with these multi-regime problems and the multiscale numerical method is required. On this point, Xu and Huang proposed the unified gas-kinetic scheme (UGKS) \cite{Xu2010A} for gas flows in all flow regimes. UGKS is presented in the finite volume framework, and is also solving the kinetic equation based on the discrete particle velocity space just as the conventional DVM. What distinguishes UGKS from the conventional DVM is that UGKS couples the particle transportation with the particle collision by the analytical solution of the governing equation in the calculation of the flux at the cell interface, which means that in the transportation process of UGKS the particle is not transferring freely as what happens in the conventional DVM but transferring simultaneously with particle collision. Then when the cell Kn number (the ratio of the mean free path and the mesh size) is very low, namely the numerical scale is much larger than the kinetic scale, the velocity distribution can evolve to near-equilibrium state in a numerical time step and the scheme will yield results consistent with the Navier-Stokes (NS) equation. Hence, by considering the particle collision in the calculation of the interface flux, UGKS turns into a multiscale scheme which can adopt mesh size and time step comparable with that of the traditional NS-equation-based CFD method in the continuum flow simulation, and it is very suitable for multiscale problems \cite{Xu2015Direct} involving both continuum and rarefied flows. The discrete unified gas-kinetic scheme (DUGKS) proposed by Guo et al.~\cite{guo2013discrete,guo2015discrete} is another multiscale scheme based on the similar idea of UGKS. Instead of using the analytical solution, in DUGKS a difference scheme of the governing equation is applied at the cell interface to calculate the multiscale flux which couples the particle transportation and the particle collision. Generally speaking UGKS and DUGKS have similar multiscale mechanisms and can yield similar results in all flow regimes.

It is worth pointing out that although UGKS and DUGKS overcome the over-dissipation of the conventional DVM in the continuum regime, the complexity and the computational cost of the numerical flux at the cell interface are increased. The increased computation is mainly due to the calculation of the equilibrium state function at the interface, which has to be calculated for every velocity point in the whole discrete velocity space. In view of this, Chen et al.~\cite{chen2016simplification} presented a simplified UGKS, Yang et al.~\cite{yang2018animproved,yang2018improved,yang2019improved} proposed an improved DVM. In these methods, at the cell interface, the flux for the macroscopic variables and the flux for the microscopic variables (namely the discrete velocity distribution function) are handled separately. For the macroscopic flux, similar to what is done in UGKS, it is constructed based on the form of the analytical solution for the kinetic model equation to include the collision effect, and is calculated by a blend of the NS flux and the DVM flux. For the microscopic flux, it is calculated by the same method as the conventional DVM, which avoids the time-consuming calculation about the equilibrium state function. These methods can achieve the same accuracy as UGKS and DUGKS, while preserve the comparable efficiency with the conventional DVM.

Recently, Su et al.~\cite{su2019can} put forward a general synthetic iteration scheme (GSIS) to find steady-state solutions of the kinetic equation. In their method the discrete velocity framework is adopted and the high-order terms obtained from the moments of the discrete velocity distribution function are extracted into the macroscopic governing equations which contain Newton's law for stress and Fourier's law for heat conduction explicitly. The method can quickly get accurate steady state solutions of gas flow in all flow regimes, although it doesn't construct a multiscale flux based on the kinetic model equation as what is done in UGKS and DUGKS. Su et al.~explained the multiscale mechanism as that the macroscopic equations help to adjust the solution obtained from the microscopic discrete velocity system so that the constraint of the cell Kn number is removed just as UGKS and DUGKS. We also mention the earlier work of Pieraccini and Puppo \cite{pieraccini2012microscopically}, where they have also extracted the macroscopic moments from the microscopic variables into the macroscopic equations, which has some similarities with the present work although their main purpose is to address the convective stability restriction caused by the fastest microscopic velocity.

Here, inspired by the work of Su et al.~\cite{su2019can}, we present another idea to achieve the multiscale property in the scheme adopting DVM framework. The present method takes the finite volume DVM framework, and just similar to the simplified UGKS by Chen et al.~\cite{chen2016simplification} and the improved DVM by Yang et al.~\cite{yang2018animproved,yang2018improved,yang2019improved}, the macroscopic flux and the microscopic flux are treated separately. Unlike UGKS and DUGKS, the macroscopic flux at the interface is not calculated by the analytical solution nor the difference scheme of the governing equation, but by the reconstruction of the moments obtained from the discrete velocity distribution function at the cell center, which is relatively concise and has some similarity in multiscale mechanism with the method of Su et al.~\cite{su2019can} although macroscopic equations containing the NS constitutive relations are not used in the present work to accelerate the convergence. The method is applicable to both steady and unsteady flows in all flow regimes, and the accuracy has been verified by test cases.

The remainder of the paper is organized as follows. In Section \ref{sec:method}, firstly a DVM-based moment method which works well in the continuum regime is introduced, then the multiscale DVM is presented by modifying the DVM-based moment method in a simple way. In Section \ref{sec:numericaltest}, the accuracy and the multiscale property of the present method are testified by several test cases. In Section \ref{sec:conclusions} a summary about the present work is given.

\section{Numerical method}\label{sec:method}
In this work, the monatomic gas is considered and the BGK-type kinetic model equation \cite{bhatnagar1954model} is solved, which has the form
\begin{equation}\label{eqn:bgk}
\frac{{\partial f}}{{\partial t}}{\rm{ + }}\vec u \cdot \frac{{\partial f}}{{\partial \vec x}} = \frac{{g - f}}{\tau },
\end{equation}
where $f$ is the gas particle velocity distribution function and $\vec u$ is the particle velocity. $\tau$ is the collision time calculated as
\begin{equation}\label{eqn:tau}
\tau  = \frac{\mu}{p},
\end{equation}
where $\mu$ and $p$ are the dynamical viscosity coefficient and the pressure respectively. The equilibrium state $g$ has a form of Maxwellian distribution,
\begin{equation}
g = \rho {\left( {\frac{\lambda }{\pi }} \right)^{\frac{3}{2}}}{e^{ - \lambda {{\vec c}^2}}},
\end{equation}
or if the Shakhov model \cite{shakhov1968generalization} is used,
\begin{equation}\label{eqn:eqstate}
g^* = \rho {\left( {\frac{\lambda }{\pi }} \right)^{\frac{3}{2}}}{e^{ - \lambda {{\vec c}^2}}}\left[ {1 + \frac{{4(1 - \Pr ){\lambda ^2}\vec q \cdot \vec c}}{{5\rho }}(2\lambda {{\vec c}^2} - 5)} \right],
\end{equation}
where $\vec c$ is the peculiar velocity $\vec c = \vec u - \vec U$ and $\vec U$ is the macroscopic gas velocity, $\vec q$ is the heat flux, $\lambda $ is a variable related to the temperature $T$ by $\lambda  = 1/(2RT)$. Macroscopic variables can be obtained by taking moments of $f$,
\begin{equation}
\vec W = \int {\vec \psi fd\Xi },
\end{equation}
\begin{equation}\label{eqn:stress}
{\bf{P}} = \int {\vec c\vec cfd\Xi },
\end{equation}
\begin{equation}\label{eqn:qflux}
\vec q = \int {\frac{1}{2}\vec c{{\vec c}^2}fd\Xi } ,
\end{equation}
where $\vec W=(\rho,\rho\vec U,\rho E)^T$ is the vector of the conservative variables, $\vec \psi$ is the vector of moments $\vec \psi  = {\left( {1,\vec u,\frac{1}{2}{{\vec u}^2}} \right)^T}$, $d\Xi  = du_xdu_ydu_z$ is the velocity space element, ${\bf{P}}$ is the stress tensor. The collision term on the right-hand side of the model equation (\ref{eqn:bgk}) follows the conservation constraint,
\begin{equation}\label{eqn:conserve}
\int {\vec \psi (g - f)d\Xi }  = \vec 0.
\end{equation}

The present method is based on the finite volume framework and the integral form of the governing equation (\ref{eqn:bgk}) is concentrated,
\begin{equation}\label{eqn:bgk_intform}
\int\limits_\Omega  {\frac{{\partial f}}{{\partial t}}dV}  + \int\limits_{\partial \Omega } {\vec u \cdot \vec nfdA}  = \int\limits_\Omega  {\frac{{g - f}}{\tau }dV},
\end{equation}
where $\Omega$ is the control volume, $dV$ is the volume element, $dA$ is the surface area element and $\vec n$ is the outward normal unit vector. Taking moments of $\vec \psi$ to Eq.~(\ref{eqn:bgk_intform}) will yield the corresponding macroscopic governing equation,
\begin{equation}\label{eqn:mac_govern}
\int\limits_\Omega  {\frac{{\partial \vec W}}{{\partial t}}dV + } \int\limits_{\partial \Omega } {\vec FdA}  = \vec 0,
\end{equation}
where the flux $\vec F$ is calculated as
\begin{equation}\label{eqn:flux_int}
\vec F = \int {\vec \psi \vec u \cdot \vec n fd\Xi }.
\end{equation}
This paper focuses on the numerical scheme for Eq.~(\ref{eqn:bgk_intform}) and Eq.~(\ref{eqn:mac_govern}). In the following, the paper will first introduce a DVM-based moment method about Eq.~(\ref{eqn:bgk_intform}) and Eq.~(\ref{eqn:mac_govern}), and then modify it to a multiscale scheme based on the DVM framework. All data reconstructions involved are done by the linear reconstruction based on least square method and the Venkatakrishnan limiter \cite{venkatakrishnan1995convergence} is used. Second-order accuracy in space and first-order accuracy in time are achieved in the current work. Accordingly, the conventional DVM mentioned below refers to the method with second-order accuracy in space and first-order accuracy in time.

\subsection{Scheme I: DVM-based moment method}\label{sec:scheme1}
In this scheme, just like the conventional DVM, at the $n$th time step the distribution function $f_{i,k}^{n}$ at the cell center is discretized both in physical space (denoted by the subscript $i$) and velocity space (denoted by the subscript $k$). The main idea of this scheme is that, solving the microscopic governing equation (\ref{eqn:bgk_intform}) to obtain the non-equilibrium moments and meanwhile solving the macroscopic governing equation (\ref{eqn:mac_govern}) to update the macroscopic variables.

\subsubsection{General construction}
We here start from the update of the macroscopic variables and the macroscopic governing equation (\ref{eqn:mac_govern}) is discretized as
\begin{equation}\label{eqn:update_mac}
\vec W_i^{n + 1} = \vec W_i^n - \frac{{\Delta t}}{{{V_i}}}\sum\limits_{j \in N(i)} {{A_{ij}}\vec F_{ij}^n},
\end{equation}
where $j$ denotes the neighboring cell of cell $i$ and $N\left( i \right)$ is the set of all of the neighbors of cell $i$. The subscript $ij$ denotes the variable at the interface between cell $i$ and $j$, and $A_{ij}$ is the interface area. The marching time step $\Delta t$ is restricted by the CFL condition based on the particle velocity ${\vec u}_k$, and is calculated as
\begin{equation}\label{eqn:cfl_mic}
\Delta {t_i} = \frac{{{V_i}}}{{\mathop {\max }\limits_k \left( {\sum\limits_{j \in N(i)} {\left( {{{\vec u}_k} \cdot {{\vec n}_{ij}}{A_{ij}}{\rm{H}}[{{\vec u}_k} \cdot {{\vec n}_{ij}}]} \right)} } \right)}}{\rm{CFL}},
\end{equation}
\begin{equation}
\Delta t = \mathop {\min }\limits_i (\Delta {t_i}),
\end{equation}
where ${\rm{H}}[x]$ is the Heaviside function defined as
\begin{equation}
{\rm{H}}[x] = \left\{ \begin{array}{l}
1,\quad x \ge 0\\
0,\quad x < 0
\end{array} \right. .
\end{equation}
The CFL number is set as $0.75$ in the work of the present paper. In Eq.~(\ref{eqn:update_mac}) the flux $\vec F_{ij}^n$ at the interface $ij$ needs to be carefully treated to avoid excess numerical dissipation in the case of low cell Kn number. For the conventional DVM of first-order temporal accuracy, this flux is calculated as the moments of the distribution function at the interface,
\begin{equation}\label{eqn:fluxmac_dvm}
\vec F_{ij,{\rm{DVM}}}^n = \sum\limits_k {{{\vec \psi }_k}{{\vec u}_k} \cdot {{\vec n}_{ij}}f_{ij,k}^n\Delta {\Xi _k}} ,
\end{equation}
where the interface distribution function $f_{ij,k}^n$ is obtained directly from the reconstruction of the initial distribution function data. Such a treatment will lead to more dissipating solution in the low-cell-Kn-number case because the reconstructed distribution function $f_{ij,k}^n$ at the interface suffers from a large deviation, which is of the order of $\Delta {x^2}$ due to the interpolation error for second-order accuracy, from the equilibrium state. In the real physical situation, the deviation should be of the order of the collision time $\tau$ when the kinetic scale is much smaller than the flow characteristic scale, just as what is revealed by the Chapman-Enskog expansion \cite{chapman1990mathematical}. Larger deviation means additional viscous stress and heat flux, making the conventional DVM very dissipating when the cell size is much larger than the mean free path (see Fig.~\ref{fig:test3_cmpvec_dvm} in Section~\ref{sec:test3} for how this excess numerical dissipation influences the results). In UGKS \cite{Xu2010A,Xu2015Direct} and DUGKS \cite{guo2013discrete,guo2015discrete}, the problem is solved by evolving the interface distribution function with a CFL time step from the initial reconstructed data based on the governing equation (\ref{eqn:bgk}). After the evolution the interface distribution function will be sufficiently close to the equilibrium state in the low-cell-Kn-number case, thus there is no additional dissipation caused by the above problem. Here, inspired by the work of Su et al.~\cite{su2019can}, the paper provides another idea. Considering that the distribution function $f_{i,k}^n$ stored at the cell center can be obtained directly and is free from the data reconstruction error, and $f_{i,k}^n$ should be sufficiently close to the equilibrium state due to the collision process of BGK equation (\ref{eqn:bgk}) in the low-cell-Kn-number case, it is reasonable to think that the flux calculated from the moments of the cell-center distribution function $f_{i,k}^n$ is reliable, i.e.
\begin{equation}
{\bf{F}}_i^n = \sum\limits_k {{{\vec \psi }_k}{{\vec u}_k}f_{i,k}^n\Delta {\Xi _k}},
\end{equation}
where ${\bf{F}}_i^n$ is the flux tensor at the cell center. On the other side, directly interpolating the flux at the cell interface from the cell-center flux data may cause stability issue. Hence, here the flux is divided into the equilibrium flux {\bf{G}} and the nonequilibrium flux {\bf{H}}, i.e.
\begin{equation}
{\bf{F}} = \int {\vec \psi \vec ugd\Xi }  + \int {\vec \psi \vec u(f - g)d\Xi }  = {\bf{G}} + {\bf{H}}.
\end{equation}
Then at the interface, the flux $\vec F_{ij}^n$ is also split as
\begin{equation}\label{eqn:flux0}
\vec F_{ij}^n = \vec G_{ij}^n + \vec H_{ij}^n,
\end{equation}
where $\vec G_{ij}^n$ and $\vec H_{ij}^n$ will be calculated in different ways just like the separate handling of convection term and diffusion term in the traditional NS-equation-based scheme. It's worth noting that in the present work the construction of the macroscopic moments is similar to the method of Pieraccini and Puppo \cite{pieraccini2012microscopically} although their aim is to address the convective stability restriction, and here we show that such a treatment is closely relevant to the dissipation property of the scheme. It is also noted that the flux splitting strategy used here has some similarities with the class of methods called micro-macro decomposition \cite{xiong2015high,jang2015high,xiong2017hierarchical}. In those micro-macro decomposition methods the nonequilibrium part of the distribution function is decomposed and evolved separately through projections, while the present work applies the equilibrium-nonequilibrium splitting to the macroscopic moments without decomposition of the distribution function and can easily recover the free transport mechanism in the high-Kn-number regime (as also pointed out by Pieraccini and Puppo \cite{pieraccini2012microscopically}).

For the nonequilibrium flux tensor {\bf{H}} at the cell center, it is calculated as
\begin{equation}\label{eqn:hfluxtensor}
{\bf{H}}_i^n = \sum\limits_k {{{\vec \psi }_k}{{\vec u}_k}f_{i,k}^n\Delta {\Xi _k}}  - \int {\vec \psi \vec u \bar g_i^nd\Xi },
\end{equation}
where $\bar g_i^n$ is the equilibrium state determined from the macroscopic variables $\bar { \vec W} _i^n$ which are obtained from the cell-center distribution function $f_{i,k}^n$,
\begin{equation}\label{eqn:macvar_int}
\bar {\vec W}_i^n = \sum\limits_k {{{\vec \psi }_k}f_{i,k}^n\Delta {\Xi _k}}.
\end{equation}
The second part on the right-hand side of Eq.~(\ref{eqn:hfluxtensor}) is directly the Euler flux tensor determined by $\bar { \vec W} _i^n$. After that, through the reconstruction of the nonequilibrium flux tensor {\bf{H}}, at the interface the nonequilibrium flux $\vec H_{ij}^n$ is calculated as
\begin{equation}\label{eqn:hflux}
\vec H_{ij}^n = \frac{1}{2}\left( {{\bf{H}}_i^n + ({{\vec x}_{ij}} - {{\vec x}_i}) \cdot \nabla {\bf{H}}_i^n + {\bf{H}}_j^n + ({{\vec x}_{ij}} - {{\vec x}_j}) \cdot \nabla {\bf{H}}_j^n} \right) \cdot {\vec n_{ij}}.
\end{equation}

For the equilibrium flux, we don't calculate the tensor ${\bf{G}}$ but directly calculate the flux $\vec G_{ij}^n$ at the cell interface. The equilibrium flux $\vec G_{ij}^n$ is actually the Euler flux which should be determined from the macroscopic variables $\vec W_{ij}^n$ at the interface. Here $\vec W_{ij}^n$ is calculated by the way used in gas-kinetic scheme (GKS) \cite{xu2001gas}. First, the macroscopic variables $\vec W$ are reconstructed and the variables can be interpolated at the two sides of the interface as
\begin{equation}
\vec W_{ij}^{n, + } = \vec W_i^n + ({\vec x_{ij}} - {\vec x_i}) \cdot \nabla \vec W_i^n,
\end{equation}
\begin{equation}
\vec W_{ij}^{n, - } = \vec W_j^n + ({\vec x_{ij}} - {\vec x_j}) \cdot \nabla \vec W_j^n.
\end{equation}
Then the macroscopic variables $\vec W_{ij}^n$ for the interface $ij$ are calculated as
\begin{equation}
\vec W_{ij}^n = \int {\vec \psi \hat g_{ij}^nd\Xi },
\end{equation}
and
\begin{equation}
\hat g_{ij}^n = \left\{ {\begin{array}{*{20}{l}}
{g_{ij}^{n, + },\quad \vec u \cdot {{\vec n}_{ij}} \ge 0}\\
{g_{ij}^{n, - },\quad \vec u \cdot {{\vec n}_{ij}} < 0}
\end{array}} \right. ,
\end{equation}
where $g_{ij}^{n, + }$ and $g_{ij}^{n, - }$ are determined by $\vec W_{ij}^{n, + }$ and $\vec W_{ij}^{n, - }$ respectively. After that, to make the scheme stable, the equilibrium flux $\vec G_{ij}^n$ at the interface is calculated as a weighting of the Euler flux and the flux of kinetic flux vector splitting (KFVS) \cite{mandal1994kinetic} which is of high stability, i.e.
\begin{equation}\label{eqn:gflux0}
\vec G_{ij}^n = \frac{{\tau _{ij}^n}}{{\tau _{ij}^n + h_{ij}^n}}\int {\vec \psi \vec u \cdot \vec n\hat g_{ij}^nd\Xi }  + \frac{{h_{ij}^n}}{{\tau _{ij}^n + h_{ij}^n}}\int {\vec \psi \vec u \cdot \vec ng_{ij}^nd\Xi },
\end{equation}
where the first part on the right-hand side is corresponding to the KFVS flux (for the calculation about the moments one can see the appendix of Ref.~\cite{xu2001gas} for a quick guide) and the second part is the Euler flux determined from $\vec W_{ij}^n$. The weight factors ${\tau _{ij}^n}/ ({\tau _{ij}^n + h_{ij}^n})$ and ${h_{ij}^n}/ ({\tau _{ij}^n + h_{ij}^n})$ are constructed based on the form of DUGKS \cite{guo2013discrete,guo2015discrete}. $\tau _{ij}^n$ is the collision time determined by $\vec W_{ij}^n$. $h_{ij}^n$ is the physical local time step determined by the CFL condition based on the macroscopic characteristic velocity, i.e.
\begin{equation}\label{eqn:physlts}
h_i^n = \frac{{{V_i}}}{{\sum\limits_{j \in N(i)} {\left( {\vec U_i^n \cdot {{\vec n}_{ij}}{A_{ij}}{\rm{H}}[\vec U_i^n \cdot {{\vec n}_{ij}}]} \right)}  + a_i^n{{\rm A}_i}}}{\rm{CFL}}_{{\rm{phys}}},
\end{equation}
and
\begin{equation}
h_{ij}^n = \min (h_i^n,h_j^n),
\end{equation}
where $a_i^n$ is the acoustic velocity and ${{{\rm A}_i}}$ is the maximum cross section area of cell $i$. For ${\rm{CFL}}_{{\rm{phys}}}$, it is set as $0.5$ in the work of the present paper. Decreasing ${\rm{CFL}}_{{\rm{phys}}}$ will increase the weight of the KFVS flux, resulting in better stability but more numerical dissipation. $h_{ij}^n$ can be viewed as the characteristic time scale based on the cell size for the interface $ij$. According to the construction of the weight factors in Eq.~(\ref{eqn:gflux0}), in the low-cell-Kn-number case (continuum case) the flux $\vec G_{ij}^n$ can recover the accurate Euler flux while in the high-cell-Kn-number case (rarefied case) it can recover the KFVS flux to stabilize the scheme. For more details about these weight factors one can refer to Ref.~\cite{yuan2019conservative} and Ref.~\cite{yuan2019multi}.

Now that we have obtained both the nonequilibrium flux $\vec H_{ij}^n$ and the equilibrium flux $\vec G_{ij}^n$, the flux $\vec F_{ij}^n$ is then calculated by Eq.~(\ref{eqn:flux0}) and the macroscopic variables can be updated to $\vec W_i^{n + 1}$ through Eq.~(\ref{eqn:update_mac}). The last thing to do is the update of the distribution function at the cell center. The microscopic governing equation (\ref{eqn:bgk_intform}) is discretized as
\begin{equation}\label{eqn:bgk_disc}
\frac{{f_{i,k}^{n + 1} - f_{i,k}^n}}{{\Delta t}} + \frac{1}{{{V_i}}}\sum\limits_{j \in N\left( i \right)} {{A_{ij}}{{\vec u}_k} \cdot {{\vec n}_{ij}}f_{ij,k}^n}  = \frac{{g_{i,k}^{n + 1} - f_{i,k}^{n + 1}}}{{\tau _i^{n + 1}}},
\end{equation}
where the collision term on the right-hand side adopts an implicit form to ensure the stability, $g_{i,k}^{n + 1}$ and $\tau _i^{n + 1}$ are both calculated by the updated macroscopic variables $\vec W_i^{n + 1}$. Eq.~(\ref{eqn:bgk_disc}) is further arranged as
\begin{equation}\label{eqn:update_mic0}
f_{i,k}^{n + 1} = \frac{{\tau _i^{n + 1}}}{{\tau _i^{n + 1} + \Delta t}}f_{i,k}^n - \frac{{\tau _i^{n + 1}\Delta t}}{{\tau _i^{n + 1} + \Delta t}}\frac{1}{{{V_i}}}\sum\limits_{j \in N\left( i \right)} {{A_{ij}}{{\vec u}_k} \cdot {{\vec n}_{ij}}f_{ij,k}^n}  + \frac{{\Delta t}}{{\tau _i^{n + 1} + \Delta t}}g_{i,k}^{n + 1}.
\end{equation}
The transportation term adopts the same treatment as in the conventional DVM, and the interface distribution function $f_{ij,k}^n$ is calculated through the reconstruction of the distribution function as
\begin{equation}
f_{ij,k}^n = \left\{ \begin{array}{l}
f_{i,k}^n + ({{\vec x}_{ij}} - {{\vec x}_i})\nabla f_{i,k}^n,\quad {{\vec u}_k} \cdot {{\vec n}_{ij}} \ge 0\\
f_{j,k}^n + ({{\vec x}_{ij}} - {{\vec x}_j})\nabla f_{j,k}^n,\quad {{\vec u}_k} \cdot {{\vec n}_{ij}} < 0
\end{array} \right. .
\end{equation}
The distribution function at the cell center can be then updated to $f_{i,k}^{n + 1}$ by Eq.~(\ref{eqn:update_mic0}).

\subsubsection{Boundary condition}\label{sec:scheme1bc}
The treatment of the boundary is one notable thing for the scheme. Here we only discuss the implementation of the diffuse reflection boundary condition with full thermal accommodation \cite{Li2005Application} on the isothermal solid wall, while for other boundary conditions one can handle them in the conventional manner. On the full-diffuse-wall boundary, just as at the interior face, the microscopic flux (flux for distribution function) and the macroscopic flux are treated separately. For the microscopic flux, the treatment is just the same as in the conventional DVM and the distribution function on the boundary face is
\begin{equation}\label{eqn:fdwmic}
f_{{\rm{w}},k}^n = \left\{ \begin{array}{l}
f_{i,k}^n + ({{\vec x}_{\rm{w}}} - {{\vec x}_i})\nabla f_{i,k}^n,\quad {{\vec u}_k} \cdot {{\vec n}_{\rm{w}}} \ge 0\\
g_{{\rm{w}},k}^{n, - },\quad {{\vec u}_k} \cdot {{\vec n}_{\rm{w}}} < 0
\end{array} \right. ,
\end{equation}
where ${{\vec x}_{\rm{w}}}$ is the center of the boundary face, ${{\vec n}_{\rm{w}}}$ is the outward normal unit vector relative to the adjacent cell $i$. $g_{{\rm{w}},k}^{n, - }$ is the Maxwellian distribution for the outcoming reflected particles and is calculated as
\begin{equation}\label{eqn:fdwgdisc}
g_{{\rm{w}},k}^{n, - } = \bar \rho _{\rm{w}}^n{\left( {\frac{{{\lambda _{\rm{w}}}}}{\pi }} \right)^{\frac{3}{2}}}{e^{ - {\lambda _{\rm{w}}}{{({{\vec u}_k} - {{\vec U}_{\rm{w}}})}^2}}},
\end{equation}
in which ${\vec U}_{\rm{w}}$ is the wall velocity and $\lambda _{\rm{w}}$ is obtained from the wall temperature. $\bar \rho _{\rm{w}}^n$ can be solved from the no-penetration condition
\begin{equation}
\sum\limits_k {{{\vec u}_k} \cdot {{\vec n}_{\rm{w}}}f_{{\rm{w}},k}^n = 0}.
\end{equation}
Then $f_{{\rm{w}},k}^n$ on the boundary face is completely determined and can be used to calculate the microscopic flux in Eq.~(\ref{eqn:update_mic0}) to update the distribution function. For the macroscopic flux which can be calculated as
\begin{equation}\label{eqn:fdwmac}
\vec F_{\rm{w}}^{n}=\vec F_{\rm{w}}^{n, + }+\vec F_{\rm{w}}^{n, - },
\end{equation}
the incoming flux $\vec F_{\rm{w}}^{n, + }$ to the wall is divided into the equilibrium flux and the nonequilibrium flux just as what we have done at the interior face to avoid excess numerical dissipation in the low-cell-Kn-number case, i.e.
\begin{equation}
\vec F_{\rm{w}}^{n, + } = \vec G_{\rm{w}}^{n, + } + \vec H_{\rm{w}}^{n, + }.
\end{equation}
The nonequilibrium flux $\vec H_{\rm{w}}^{n, + }$ is calculated directly from the variables at the center of the adjacent cell $i$ as
\begin{equation}
\vec H_{\rm{w}}^{n, + } = \sum\limits_k^{{{\vec u}_k} \cdot {{\vec n}_{\rm{w}}} \ge 0} {{{\vec \psi }_k}{{\vec u}_k} \cdot {{\vec n}_{\rm{w}}}f_{i,k}^n\Delta {\Xi _k}}  - \int_{\vec u \cdot {{\vec n}_{\rm{w}}} \ge 0} {\vec \psi \vec u \cdot {{\vec n}_{\rm{w}}}\bar g_i^nd\Xi },
\end{equation}
which means the first-order accuracy for this part of flux. The equilibrium flux $\vec G_{\rm{w}}^{n, + }$ is calculated by
\begin{equation}
\vec G_{\rm{w}}^{n, + } = \int_{\vec u \cdot {{\vec n}_{\rm{w}}} \ge 0} {\vec \psi \vec u \cdot {{\vec n}_{\rm{w}}}g_{\rm{w}}^{n, + }d\Xi },
\end{equation}
where ${g_{\rm{w}}^{n, + }}$ is determined by the macroscopic variables $\vec W_{\rm{w}}^{n, + }$ which are obtained through reconstruction with second-order accuracy as
\begin{equation}
\vec W_{\rm{w}}^{n, + } = \vec W_i^n + ({\vec x_{\rm{w}}} - {\vec x_i}) \cdot \nabla \vec W_i^n.
\end{equation}
Such a construction makes the main part of $\vec F_{\rm{w}}^{n, + }$ second-order accurate in the continuum flow case. After $\vec F_{\rm{w}}^{n, + }$ is obtained, the outcoming flux $\vec F_{\rm{w}}^{n, - }$ is calculated as
\begin{equation}
\vec F_{\rm{w}}^{n, - } = \int_{\vec u \cdot {{\vec n}_{\rm{w}}} < 0} {\vec \psi \vec u \cdot {{\vec n}_{\rm{w}}}g_{\rm{w}}^{n, - }d\Xi },
\end{equation}
where the reflected Maxwellian distribution $g_{\rm{w}}^{n, - }$ is
\begin{equation}
g_{\rm{w}}^{n, - } = \rho _{\rm{w}}^n{\left( {\frac{{{\lambda _{\rm{w}}}}}{\pi }} \right)^{\frac{3}{2}}}{e^{ - {\lambda _{\rm{w}}}{{({{\vec u}_k} - {{\vec U}_{\rm{w}}})}^2}}},
\end{equation}
which is different from the discrete $g_{{\rm{w}},k}^{n, - }$ obtained by Eq.~(\ref{eqn:fdwgdisc}) because here $\rho _{\rm{w}}^n$ should be constrained by the no-penetration condition for the macroscopic flux
\begin{equation}
F_{{\rm{w,}}\rho }^{n, + } + F_{{\rm{w,}}\rho }^{n, - } = 0,
\end{equation}
where $F_{{\rm{w,}}\rho }^{n, + }$ and $F_{{\rm{w,}}\rho }^{n, - }$ are the incoming mass flux and the outcoming mass flux in $\vec F_{\rm{w}}^{n, + }$ and $\vec F_{\rm{w}}^{n, - }$ respectively. The macroscopic flux $\vec F_{\rm{w}}^{n}$ is then completely determined and can be used in the update of the macroscopic variables Eq.~(\ref{eqn:update_mac}).

\subsubsection{Artificial dissipation}
When applying the above scheme to the supersonic flow simulation with high Re number, artificial dissipation should be added to suppress the oscillation induced by the shock wave. The artificial dissipation is introduced by the following two measures. First, the weight of the KFVS flux for $\vec G_{ij}^n$ in Eq.~(\ref{eqn:gflux0}) is increased at the discontinuity region by calculating the collision time $\tau _{ij}^n$ as
\begin{equation}\label{eqn:artau}
\tau _{ij}^n = \tau _{ij,{\rm{physical}}}^n + \tau _{ij,{\rm{artificial}}}^n = \frac{{\mu _{ij}^n}}{{p_{ij}^n}} + \frac{{\left| {p_{ij}^{n, + } - p_{ij}^{n, - }} \right|}}{{\left| {p_{ij}^{n, + } + p_{ij}^{n, - }} \right|}}{h_{ij}},
\end{equation}
where $p_{ij}^{n, + }$ and $p_{ij}^{n, - }$ are pressures calculated from the reconstructed macroscopic variables $\vec W_{ij}^{n, + }$ and $\vec W_{ij}^{n, - }$ at the two sides of the interface. By applying Eq.~(\ref{eqn:artau}), the pressure discontinuity will increase the weight of the KFVS flux and the oscillation induced by such discontinuity can be suppressed. Second, an artificial viscous term is introduced explicitly into the interface flux $\vec F_{ij}^n$ by amplifying the nonequilibrium flux $\vec H_{ij}^n$, that is
\begin{equation}\label{eqn:artvis}
\vec F_{ij}^n = \vec G_{ij}^n + (1 + \frac{{h_{ij}^n}}{{\tau _{ij}^n + h_{ij}^n}}\frac{{\tau _{ij,{\rm{artificial}}}^n}}{{\tau _{ij,{\rm{physical}}}^n}})\vec H_{ij}^n,
\end{equation}
where the factor ${h_{ij}^n}/({\tau _{ij}^n + h_{ij}^n})$ is used to switch off the artificial viscosity in the case of rarefied flow.

\subsubsection{Remarks}\label{sec:scheme1re}
At the end of this part, we make some remarks about the above scheme. By interpolating the nonequilibrium flux from the cell center to the cell interface, the scheme avoids the excessive dissipation suffered by the conventional DVM in the continuum flow simulation. The scheme adopts the DVM framework, but the microscopic flux (the flux of the distribution function) in Eq.~(\ref{eqn:update_mic0}) and the macroscopic flux calculated by Eq.~(\ref{eqn:flux0}) or Eq.~(\ref{eqn:artvis}) are not consistent, so the macroscopic variables $\vec W_i^n$ are not equal to the macroscopic variables $\bar {\vec W}_i^n$ calculated by Eq.~(\ref{eqn:macvar_int}) from the integration of the distribution function at the cell center. The discrete distribution function $f_{i,k}^n$ is only used to obtain the nonequilibrium flux $\vec H_{ij}^n$ and the method is essentially a moment method. In the rarefied flow simulation with a high cell Kn number, the relaxation (or collision) process of the microscopic governing equation drives the distribution function $f_{i,k}^n$ to the equilibrium state determined by the macroscopic variables very slowly (see Eq.~(\ref{eqn:update_mic0}) for reference), making the distribution function $f_{i,k}^n$ and the nonequilibrium flux $\vec H_{ij}^n$ nearly decoupled with the macroscopic variables $\vec W_i^n$. Due to this decoupling, in the rarefied flow simulation, the macroscopic variables $\vec W_i^n$ may present a fake result which is spoiled by the discretization error. Our test cases show that the scheme I works well in the continuum flow cases but works improperly in the rarefied flow cases. One recipe (not applied in the current work) for this problem is to view the macroscopic variables $\bar {\vec W}_i^n$ obtained from $f_{i,k}^n$ by Eq.~(\ref{eqn:macvar_int}) as the real solution of the scheme. This strategy is useful in the single-scale single-regime simulation, but may not work well in the multiscale multi-regime simulation because the unreliable variables $\vec W_i^n$ in the rarefied region can still spoil the solution of the adjacent (either in time or in space) continuum region.

\subsection{Scheme II: Multiscale DVM}\label{sec:scheme2}
In the scheme I, the macroscopic variables $\vec W_i^n$ and the distribution function $f_{i,k}^n$ get decoupled when the cell Kn number is high and the scheme becomes unreliable in such a case. Here a very concise treatment can be used to fix this problem. Just as mentioned in the end of Section \ref{sec:scheme1}, the macroscopic variables $\bar {\vec W}_i^n$ obtained from the integration of $f_{i,k}^n$ are more reliable. Therefore, after the update of the distribution function $f_{i,k}^{n+1}$ through Eq.~(\ref{eqn:update_mic0}), the macroscopic variables $\vec W_i^{n+1}$ at the next time step are recalculated by integrating $f_{i,k}^{n+1}$ as
\begin{equation}\label{eqn:update_mac1}
\vec W_i^{n + 1} = \sum\limits_k {{{\vec \psi }_k}f_{i,k}^{n + 1}\Delta {\Xi _k}}.
\end{equation}
This means that the macroscopic variable updated through the macroscopic flux ${\vec F_{ij}^n}$ by Eq.~(\ref{eqn:update_mac}) is only an intermediate to get the equilibrium state in Eq.~(\ref{eqn:update_mic0}), and to avoid ambiguousness these equations are rewritten in the scheme II as
\begin{equation}\label{eqn:update_macin}
\tilde {\vec W}_i^{n + 1} = \vec W_i^n - \frac{{\Delta t}}{{{V_i}}}\sum\limits_{j \in N(i)} {{A_{ij}}\vec F_{ij}^n},
\end{equation}
\begin{equation}\label{eqn:update_mic1}
f_{i,k}^{n + 1} = \frac{{\tilde \tau _i^{n + 1}}}{{\tilde \tau _i^{n + 1} + \Delta t}}f_{i,k}^n - \frac{{\tilde \tau _i^{n + 1}\Delta t}}{{\tilde \tau _i^{n + 1} + \Delta t}}\frac{1}{{{V_i}}}\sum\limits_{j \in N\left( i \right)} {{A_{ij}}{{\vec u}_k} \cdot {{\vec n}_{ij}}f_{ij,k}^n}  + \frac{{\Delta t}}{{\tilde \tau _i^{n + 1} + \Delta t}}\tilde g_{i,k}^{n + 1},
\end{equation}
where $\tilde {\vec W}_i^{n + 1}$ are the intermediate macroscopic variables and $\tilde g_{i,k}^{n + 1}$ is the intermediate equilibrium state determined by $\tilde {\vec W}_i^{n + 1}$. By simply applying Eq.~(\ref{eqn:update_mac1}), the macroscopic variables $\vec W_i^n$ and the distribution function $f_{i,k}^n$ are closely coupled and the scheme turns into a multiscale scheme for all flow regimes. Further investigation for the mechanism will ensue. Combining Eq.~(\ref{eqn:update_mac1}), Eq.~(\ref{eqn:update_macin}) and Eq.~(\ref{eqn:update_mic1}) will yield
\begin{equation}
\vec W_i^{n + 1} = \frac{{\tilde \tau _i^{n + 1}}}{{\tilde \tau _i^{n + 1} + \Delta t}}\vec W_i^n - \frac{{\tilde \tau _i^{n + 1}\Delta t}}{{\tilde \tau _i^{n + 1} + \Delta t}}\frac{1}{{{V_i}}}\sum\limits_{j \in N\left( i \right)} {{A_{ij}}\vec F_{ij,{\rm{DVM}}}^n}  + \frac{{\Delta t}}{{\tilde \tau _i^{n + 1} + \Delta t}}(\vec W_i^n - \frac{{\Delta t}}{{{V_i}}}\sum\limits_{j \in N\left( i \right)} {{A_{ij}}\vec F_{ij}^n} ),
\end{equation}
and it can be arranged as
\begin{equation}\label{eqn:update_mac2}
\vec W_i^{n + 1} = \vec W_i^n - \frac{{\Delta t}}{{{V_i}}}\sum\limits_{j \in N\left( i \right)} {{A_{ij}}\left( {\frac{{\tilde \tau _i^{n + 1}}}{{\tilde \tau _i^{n + 1} + \Delta t}}\vec F_{ij,{\rm{DVM}}}^n + \frac{{\Delta t}}{{\tilde \tau _i^{n + 1} + \Delta t}}\vec F_{ij}^n} \right)} .
\end{equation}
Eq.~(\ref{eqn:update_mac2}) shows that the macroscopic variables $\vec W_{i}^{n + 1}$ are intrinsically updated through the weighting of $\vec F_{ij,{\rm{DVM}}}^n$ and $\vec F_{ij}^n$, where the former is the conventional DVM flux calculated by Eq.~(\ref{eqn:fluxmac_dvm}) and the latter is the flux calculated by Eq.~(\ref{eqn:artvis}) which is of relatively low numerical dissipation in the low-cell-Kn-number case. Such a construction is quite similar to other multiscale methods like UGKS \cite{Xu2010A,Xu2015Direct} and DUGKS \cite{guo2013discrete,guo2015discrete}. In those methods the flux at the cell interface couples the particle free transportation with the particle collision through the BGK equation, and can be also approximately viewed as the weighting of the conventional DVM flux and the flux based on NS equation. Furthermore, the weight factors for the fluxes in Eq.~(\ref{eqn:update_mac2}) are very similar to those in DUGKS \cite{guo2013discrete,guo2015discrete}, which are deduced from the discretization of the BGK equation likewise. Thus, the scheme applying Eq.~(\ref{eqn:update_mac1}) can get the multiscale property based on the same mechanism as UGKS and DUGKS.

It is worth noting that, the scheme using Eq.~(\ref{eqn:update_mac1}) works well in the uniform mesh, but in the non-uniform mesh due to the constraint of the global CFL condition the marching time step ${\Delta t}$ can be very small and the weight for the conventional DVM flux $\vec F_{ij,{\rm{DVM}}}^n$ will be overestimated as shown in Eq.~(\ref{eqn:update_mac2}), leading to excessive dissipation. In view of this, in the scheme II the macroscopic variables are not directly updated by Eq.~(\ref{eqn:update_mac1}), but by modifying Eq.~(\ref{eqn:update_mac2}) as
\begin{equation}\label{eqn:update_mac3}
\vec W_i^{n + 1} = \vec W_i^n - \frac{{\Delta t}}{{{V_i}}}\sum\limits_{j \in N\left( i \right)} {{A_{ij}}\left( {\frac{{\tilde \tau _i^{n + 1}}}{{\tilde \tau _i^{n + 1} + h_i^n}}\vec F_{ij,{\rm{DVM}}}^n + \frac{{h_i^n}}{{\tilde \tau _i^{n + 1} + h_i^n}}\vec F_{ij}^n} \right)} ,
\end{equation}
which can be further simplified as
\begin{equation}\label{eqn:update_mac3s}
\vec W_i^{n + 1} = \vec W_i^n + \frac{{\tilde \tau _i^{n + 1}}}{{\tilde \tau _i^{n + 1} + h_i^n}}\Delta \vec W_{i,{\rm{DVM}}}^{n + 1} + \frac{{h_i^n}}{{\tilde \tau _i^{n + 1} + h_i^n}}\Delta \tilde {\vec W}_i^{n + 1},
\end{equation}
where $h_i^n$ is the physical local time step obtained from Eq.~(\ref{eqn:physlts}). In addition, considering that in Eq.~(\ref{eqn:update_mac3}) (and also in Eq.~(\ref{eqn:update_mac2})) there is already a weighting process which introduces the dissipative conventional DVM flux into the flux at the interface, the calculation of the equilibrium flux $\vec G_{ij}^n$ (i.e. Eq.~(\ref{eqn:gflux0})) is modified in the scheme II as
\begin{equation}\label{eqn:gflux1}
\vec G_{ij}^n = \frac{{\tau _{ij,{\rm{artificial}}}^n}}{{\tau _{ij,{\rm{artificial}}}^n + h_{ij}^n}}\int {\vec \psi \vec u \cdot \vec n\hat g_{ij}^nd\Xi }  + \frac{{h_{ij}^n}}{{\tau _{ij,{\rm{artificial}}}^n + h_{ij}^n}}\int {\vec \psi \vec u \cdot \vec ng_{ij}^nd\Xi },
\end{equation}
namely, the physical collision time $\tau _{ij,{\rm{physical}}}^n$ in $\tau _{ij}^n$ (see Eq.~(\ref{eqn:artau}) for reference) is removed from the weight factors of Eq.~(\ref{eqn:gflux0}) to reduce the weight of the KFVS flux in $\vec G_{ij}^n$ and avoid overmuch numerical dissipation.

To make the scheme II more clear, the computation procedure in one step is listed as follows:
\begin{description}
    \item[Step 1.] Calculate the nonequilibrium flux tensor ${\bf{H}}_i^n$ at the cell center by Eq.~(\ref{eqn:hfluxtensor}), and then calculate the nonequilibrium flux $\vec H_{ij}^n$ at the interface by Eq.~(\ref{eqn:hflux}) through data reconstruction.
    \item[Step 2.] Calculate the equilibrium flux $\vec G_{ij}^n$ at the cell interface by Eq.~(\ref{eqn:gflux1}) with data reconstruction.
    \item[Step 3.] Calculate the macroscopic flux $\vec F_{ij}^n$ at the interface by Eq.~(\ref{eqn:artvis}), where $\tau _{ij}^n$ is calculated by Eq.~(\ref{eqn:artau}).
    \item[Step 4.] Calculate the intermediate macroscopic variables $\tilde {\vec W}_i^{n + 1}$ by Eq.~(\ref{eqn:update_macin}).
    \item[Step 5.] Update the distribution function $f_{i,k}^{n + 1}$ by Eq.~(\ref{eqn:update_mic1}), during which the reconstruction of the distribution function is implemented.
    \item[Step 6.] Update the macroscopic variables $\vec W_i^{n + 1}$ by Eq.~(\ref{eqn:update_mac3s}).
\end{description}

Finally, we'd like to point out that in Eq.~(\ref{eqn:update_mac3}), because the weight factors for the fluxes are calculated by the cell-related variables ${\tilde \tau _i^{n + 1}}$ and ${h_i^n}$, the actual macroscopic flux across the cell interface will have different values for the two cells on the two sides, which will cause conservation issue. Fortunately, the fluxes ${\vec F_{ij,{\rm{DVM}}}^n}$ and ${\vec F_{ij}^n}$ satisfy the conservation constraint separately (i.e. the same for cells on both sides), so this conservation issue is not so serious. For simulations which have initial-value dependence, some simple tricks (such as the mass compensation tricks used in \cite{Zhu2016Implicit} and \cite{yuan2019conservative}) can be applied to fix the problem. Also, one can further modify Eq.~(\ref{eqn:update_mac3}) and calculate the weight factors using the interface-related variables such as $\tau_{ij}^n$ and ${h_{ij}^n}$, which solves this issue completely with increased computational complexity and computational cost. In the present work, we directly apply Eq.~(\ref{eqn:update_mac3s}) without any additional tricks, and the numerical results seem not spoiled by the conservation issue.

\section{Numerical results}\label{sec:numericaltest}
To investigate the performance of the scheme II (the multiscale DVM) as well as the scheme I, three sets of test cases are carried out in this section. The cases include the Sod's shock tube, the lid-driven cavity flow and the laminar boundary layer flow over a flat plate, covering a variety of conditions such as unsteady, steady, continuum, rarefied. In all of the test cases the working gas is monatomic and the Shakhov model Eq.~(\ref{eqn:eqstate}) with a Prandtl number equal to $2/3$ is used.

\subsection{Sod's shock tube}\label{sec:test1}
The 1D test case of Sod's shock tube is performed to show the capability of the schemes in unsteady flow simulation for various flow regimes. The computational domain is $x \in [0,1]$ with a length $L=1$ and the initial field is
\begin{equation}
\vec W = \left\{ \begin{array}{l}
{(1,0,1)^T},x \le 0.5\\
{(0.125,0,0.1)^T},x > 0.5
\end{array} \right.
\end{equation}
with a jump at the center of the field. The computational domain is discretized into 100 uniform cells. For the particle velocity space, the discrete domain is set as $u \in [ - 8,8]$ and it is also discretized uniformly into 100 cells. The hard sphere (HS) model is used with heat index $\omega=0.5$. Three Kn numbers, which are defined by the length of the physical-space domain $L$, are considered, i.e. $1.227\times10^{-5}$, $1.227\times10^{-3}$ and $1.227$. Because of the initial discontinuity in the field, the Venkatakrishnan limiter \cite{venkatakrishnan1995convergence} is used. The time step is set as $0.001$ and the result at $t = 0.15$ is investigated.

\begin{figure}
\centering
\subfigure[Density]{\includegraphics[width=0.48\textwidth]{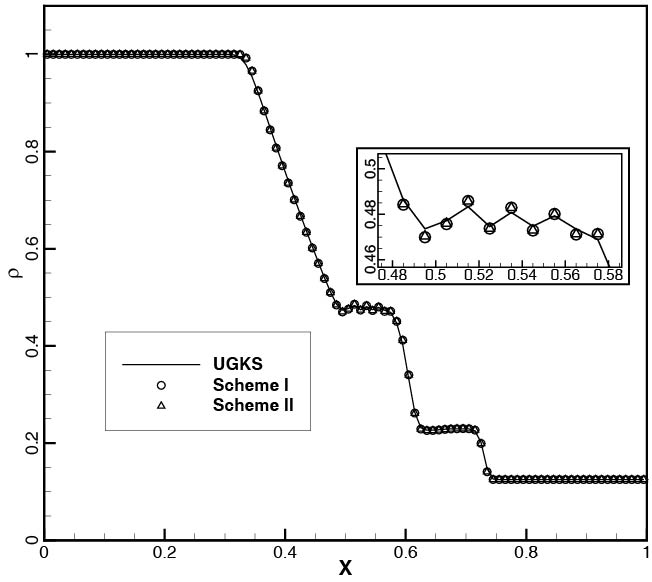}}\hspace{0.02\textwidth}%
\subfigure[Velocity]{\includegraphics[width=0.48\textwidth]{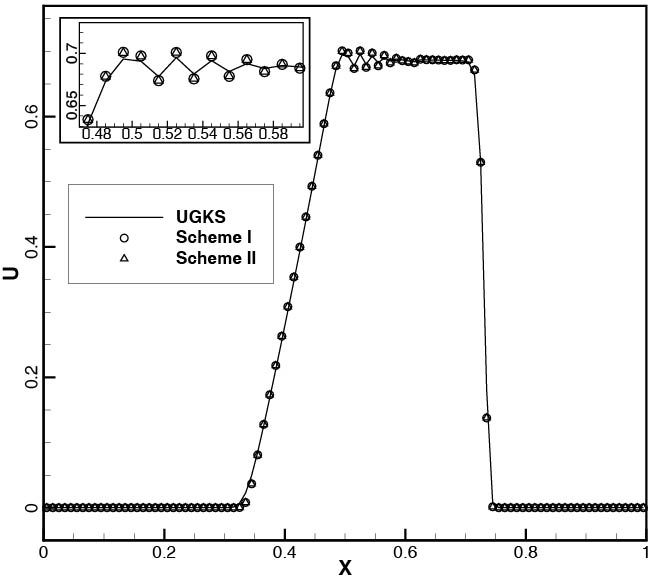}}\\
\subfigure[Temperature ($1/(2\lambda) = RT$)]{\includegraphics[width=0.48\textwidth]{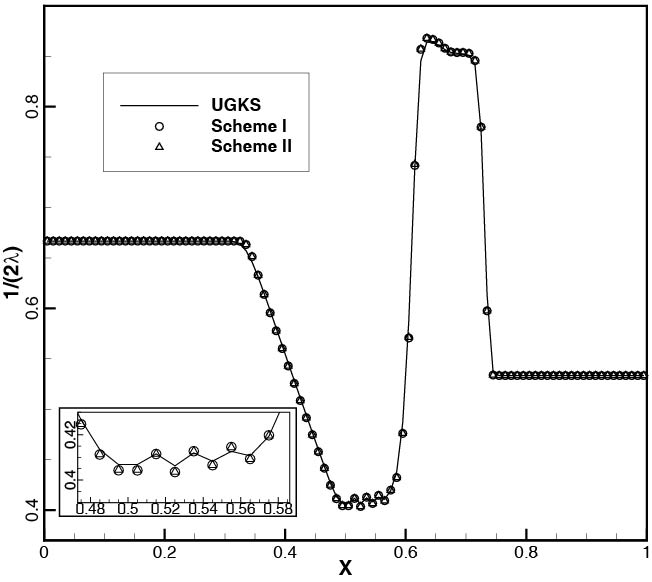}}
\caption{\label{fig:test1_kn-5}Sod's shock tube at Kn = $1.227\times10^{-5}$. }
\end{figure}

\begin{figure}
\centering
\subfigure[Density]{\includegraphics[width=0.48\textwidth]{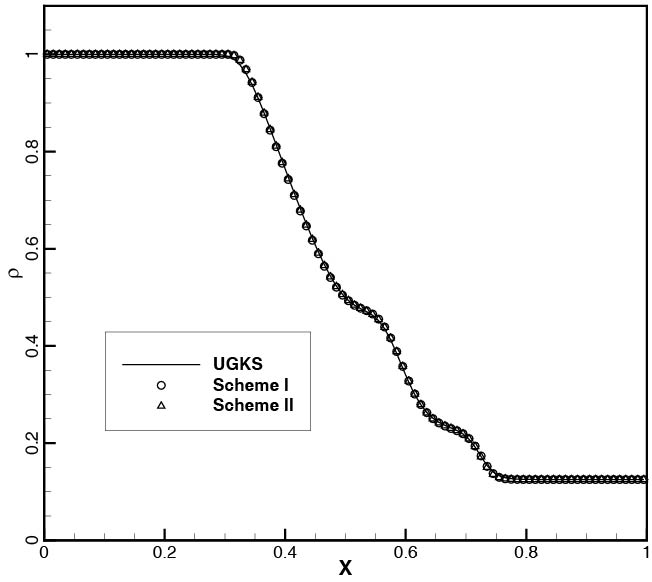}}\hspace{0.02\textwidth}%
\subfigure[Velocity]{\includegraphics[width=0.48\textwidth]{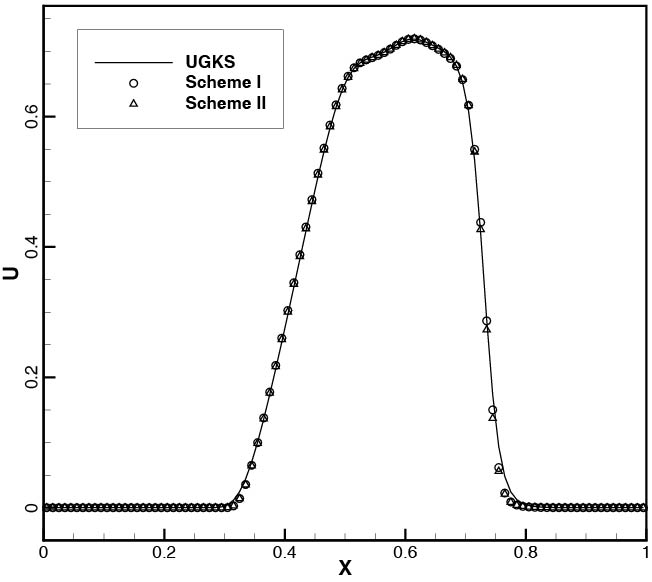}}\\
\subfigure[Temperature ($1/(2\lambda) = RT$)]{\includegraphics[width=0.48\textwidth]{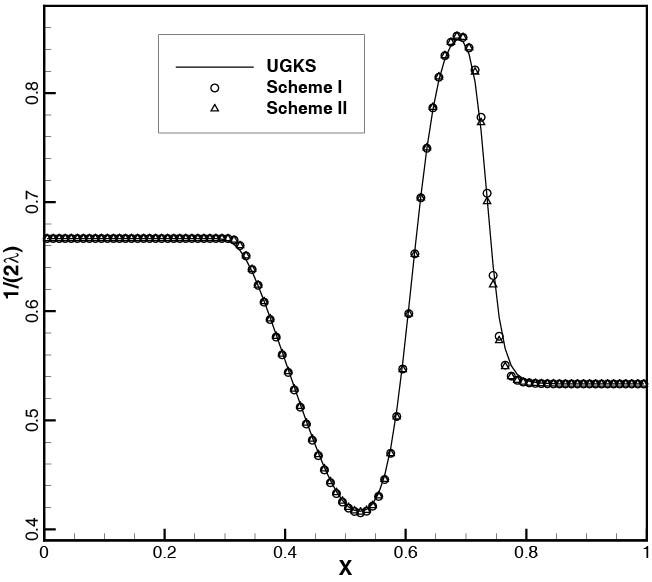}}
\caption{\label{fig:test1_kn-3}Sod's shock tube at Kn = $1.227\times10^{-3}$. }
\end{figure}

\begin{figure}
\centering
\subfigure[Density]{\includegraphics[width=0.48\textwidth]{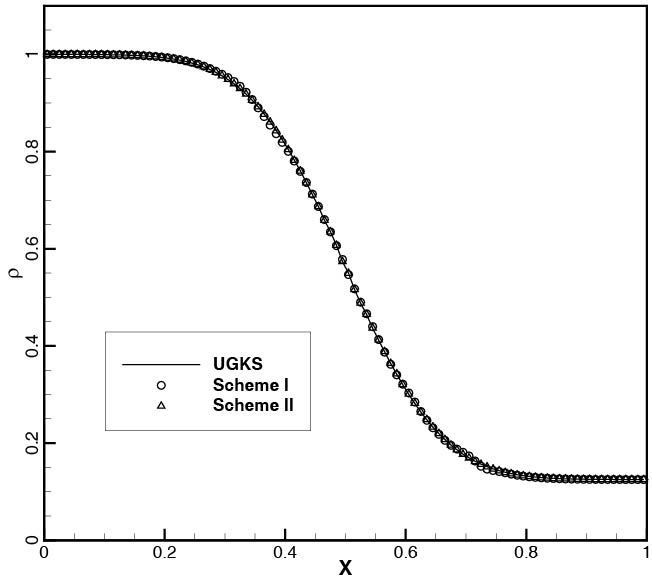}}\hspace{0.02\textwidth}%
\subfigure[Velocity]{\includegraphics[width=0.48\textwidth]{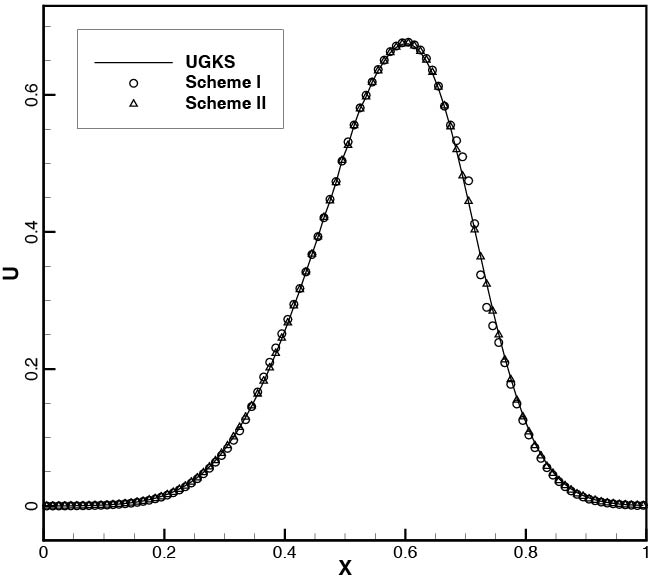}}\\
\subfigure[Temperature ($1/(2\lambda) = RT$)]{\includegraphics[width=0.48\textwidth]{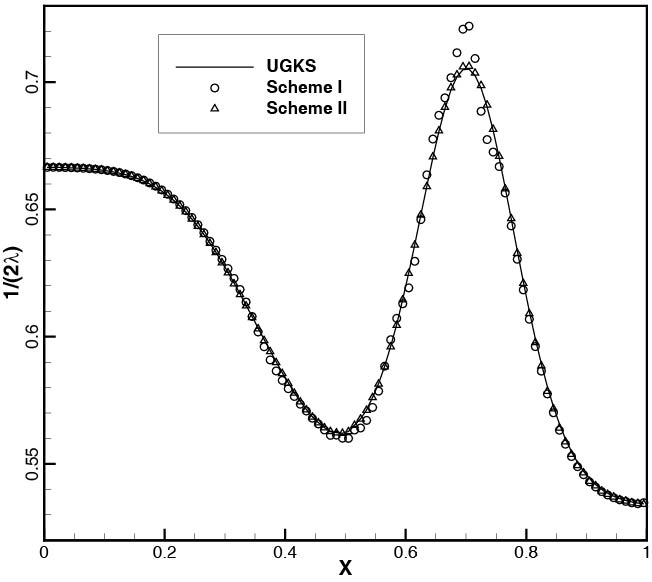}}
\caption{\label{fig:test1_kn1}Sod's shock tube at Kn = $1.227$. }
\end{figure}

The results are shown in Fig.~\ref{fig:test1_kn-5}, Fig.~\ref{fig:test1_kn-3} and Fig.~\ref{fig:test1_kn1}. The reference results are calculated by UGKS \cite{Xu2010A,Xu2015Direct} under the same simulation setup. For the case ${\rm{Kn}}=1.227\times10^{-5}$ in the continuum regime, the three sets of results agree well, and the results obtained from the scheme I and the scheme II completely overlap. The results all suffer from some oscillations at the position of the initial discontinuity. These oscillations can be suppressed by varieties of measures, such as decreasing ${\rm{CFL}}_{{\rm{phys}}}$ in Eq.~(\ref{eqn:physlts}), adding more artificial dissipation, modifying the reconstruction method, but here we don't pay much effort on this. For the case ${\rm{Kn}}=1.227\times10^{-3}$, the three sets of results are also in good consistence. But comparing with the results of ${\rm{Kn}}=1.227\times10^{-5}$, slight differences exist at the position of the shock wave between the results from scheme I and the results from scheme II. For the case ${\rm{Kn}}=1.227$, the results from UGKS and scheme II agree well while the results from scheme I have some observable errors, which implies the invalidation of the scheme I in such rarefied case just as discussed in Section~\ref{sec:scheme1re}. Thus, it can be seen that the multiscale DVM (the scheme II) works well in this set of test cases for various flow regimes, while the scheme I is only valid for the continuum flow regime.

\clearpage

\subsection{Lid-driven cavity flow}
\begin{figure}
\centering
\subfigure[Nonuniform $61\times61$ physical-space mesh\label{fig:test2_re1000_mesh}]{\includegraphics[width=0.48\textwidth]{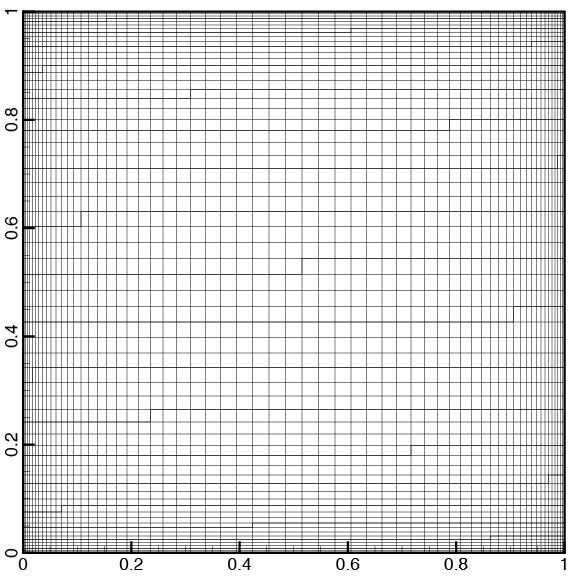}}\hspace{0.02\textwidth}%
\subfigure[Streamlines (scheme II)]{\includegraphics[width=0.48\textwidth]{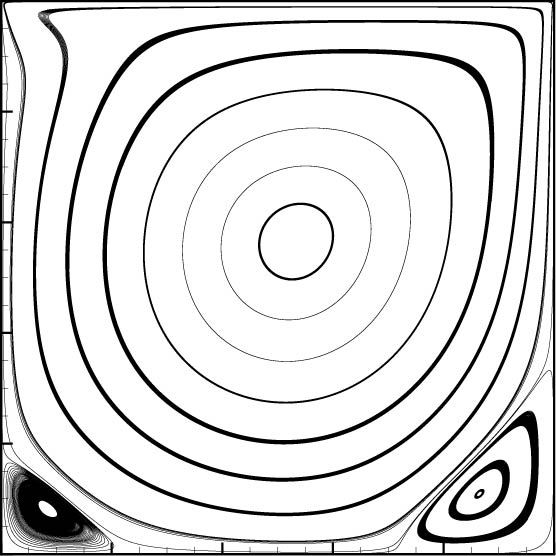}}\\
\subfigure[$U_y$ along the horizontal central line and $U_x$ along the vertical central line]{\includegraphics[width=0.6\textwidth]{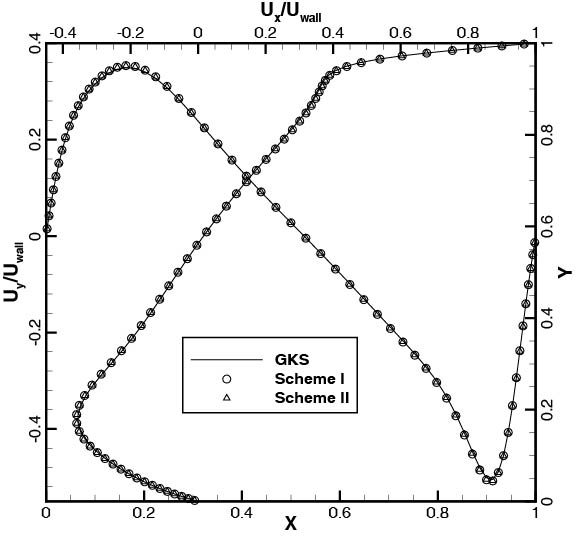}}
\caption{\label{fig:test2_re1000}Cavity flow at ${\rm{Re}}=1000$.}
\end{figure}

The 2D test case of lid-driven cavity flow is conducted to research whether the schemes can accurately simulate the viscous effect in various flow regimes. In this test case the gas flow inside a closed square cavity is driven by the tangentially moving lid. The Mach number defined by the lid velocity $U_{\rm{wall}}$ and the acoustic velocity ${a_{{\rm{wall}}}}$ is around $0.16$, where ${a_{{\rm{wall}}}}$ is defined by the fixed wall temperature $T_{\rm{wall}}$. Aiming to investigate the property of the schemes from continuum to rarefied regimes, three degrees of viscosity (or degrees of rarefaction), including ${\rm{Re}}=1000$ and ${\rm{Kn}}=0.075, 10$, are considered, where the reference length is the cavity width $L$. The HS molecular model is applied and the heat index is $\omega=0.5$. The physical-space computational domain is discretized into $61\times61$ structured mesh. For the case ${\rm{Re}}=1000$ a nonuniform mesh is used with a minimum mesh size $0.004L$ near the walls (as shown in Fig.~\ref{fig:test2_re1000_mesh}), and for the cases ${\rm{Kn}}=0.075, 10$ a uniform mesh is used. The discrete velocity-space domain is set as a square $\left\{ {{{({u_x},{u_y})}^T}|{u_x} \in [ - 6{a_{{\rm{wall}}}},6{a_{{\rm{wall}}}}],{u_y} \in [ - 6{a_{{\rm{wall}}}},6{a_{{\rm{wall}}}}]} \right\}$, and is discretized into $20 \times 20$, $50 \times 50$, $120 \times 120$ uniform cells for the cases ${\rm{Re}}=1000$ and ${\rm{Kn}}=0.075, 10$ respectively. On the walls of the cavity, the diffuse reflection boundary condition with full thermal accommodation is imposed and the implementation of this boundary condition is discussed in Section~\ref{sec:scheme1bc}.

\begin{figure}
\centering
\subfigure[Temperature contours (color band: UGKS, solid line: scheme I, dashed line: scheme II)]{\includegraphics[width=0.48\textwidth]{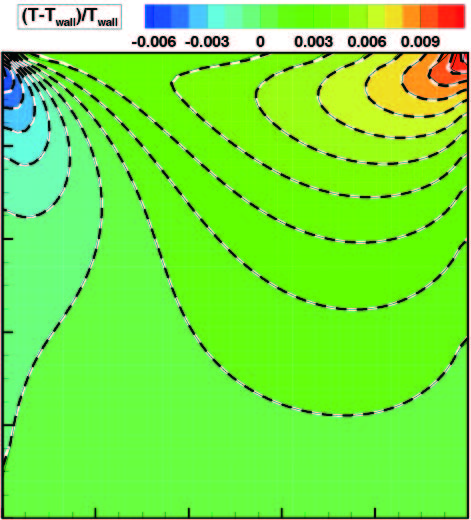}}\hspace{0.02\textwidth}%
\subfigure[Heat flux (dot: UGKS, line: scheme II)]{\includegraphics[width=0.48\textwidth]{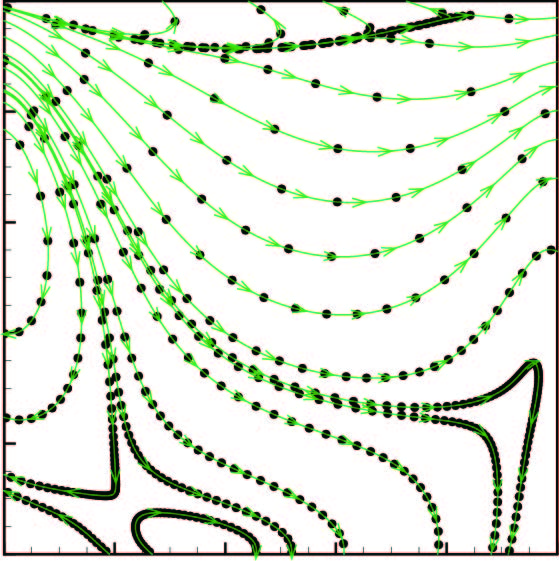}}\\
\subfigure[$U_y$ along the horizontal central line and $U_x$ along the vertical central line]{\includegraphics[width=0.6\textwidth]{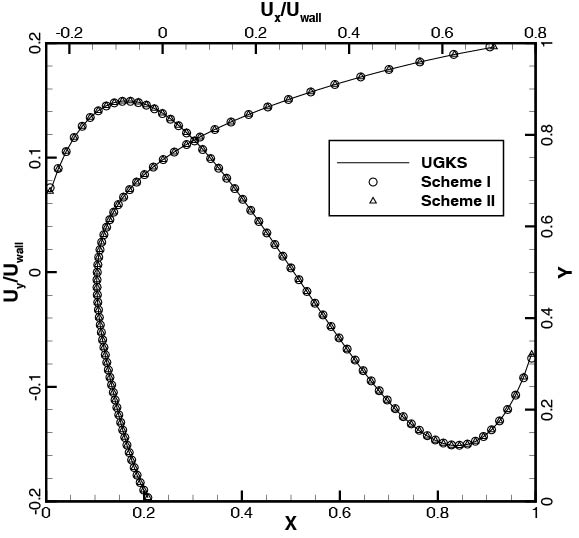}}
\caption{\label{fig:test2_kn75}Cavity flow at ${\rm{Kn}}=0.075$.}
\end{figure}

\begin{figure}
\centering
\subfigure[Temperature contours (color band: UGKS, solid line: scheme I, dashed line: scheme II)]{\includegraphics[width=0.48\textwidth]{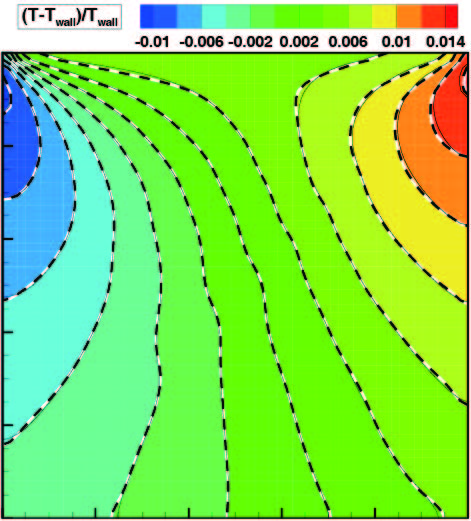}}\hspace{0.02\textwidth}%
\subfigure[Heat flux (dot: UGKS, line: scheme II)]{\includegraphics[width=0.48\textwidth]{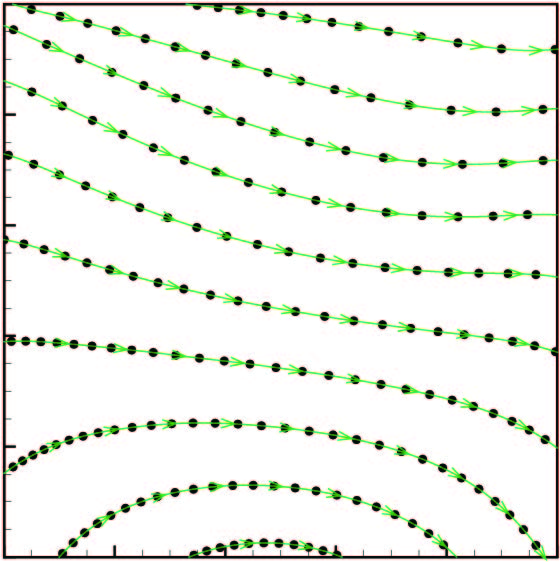}}\\
\subfigure[$U_y$ along the horizontal central line and $U_x$ along the vertical central line]{\includegraphics[width=0.6\textwidth]{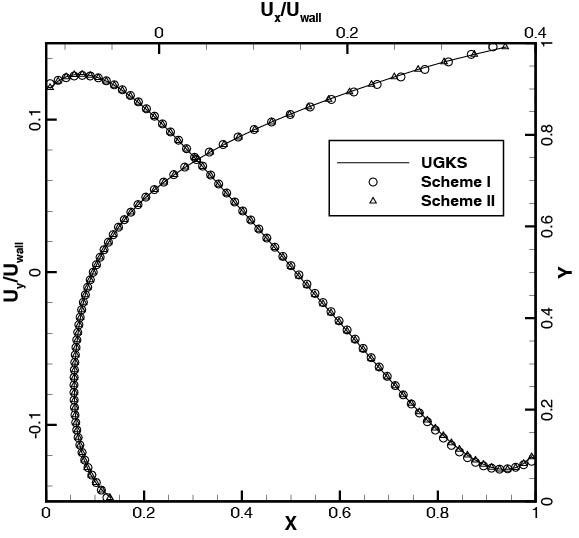}}
\caption{\label{fig:test2_kn10}Cavity flow at ${\rm{Kn}}=10$.}
\end{figure}

The results are shown in Fig.~\ref{fig:test2_re1000}, Fig.~\ref{fig:test2_kn75} and Fig.~\ref{fig:test2_kn10}. For the continuum case ${\rm{Re}}=1000$, the results from scheme I and scheme II overlap with each other, and they agree well with the results obtained from GKS \cite{xu2001gas}, which can yield NS solutions in second-order accuracy, under the same simulation setup. For the case ${\rm{Kn}}=0.075$, generally speaking the three sets of results agree well. More precisely, the results from scheme II coincide with the results from UGKS perfectly, while the results from scheme I agree well with other two sets of results except some very minor and insignificant mismatches near the wall boundary. For the case ${\rm{Kn}}=10$, the results from scheme II and UGKS are in very good consistence while the results of scheme I suffer some small deviations, which are larger than the mismatches in the case ${\rm{Kn}}=0.075$, near the wall. In conclusion, for this set of test cases the multiscale DVM (the scheme II) can give results identical to the results of GKS in the continuum case and the results of UGKS in rarefied cases, meanwhile the scheme I can also give acceptable results except some small deviations in the rarefied cases.

\clearpage

\subsection{Laminar boundary layer flow over a flat plate}\label{sec:test3}
\begin{figure}
\centering
\includegraphics[width=0.7\textwidth]{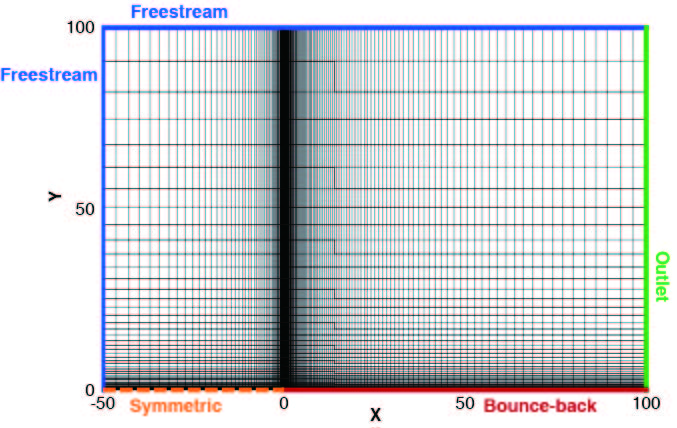}
\caption{\label{fig:test3_mesh}Mesh and boundary condition setup for the laminar boundary flow over a flat plate (case of $\Delta y_{\rm{min}}=0.1$).}
\end{figure}

For a multiscale scheme, besides describing the dynamics of particle transportation and collision when the numerical cell size is comparable with the kinetic scale, it should recover the accurate viscous hydrodynamic mechanism when the numerical scale is much larger than the kinetic scale. Here the test case of laminar boundary layer flow over a flat plate is conducted to test if the schemes can recover the accurate constitutive relations in the continuum flow regime just as an NS-equation-based scheme without excess numerical dissipation under a cell size much larger than the kinetic scale. The freestream condition is $\rm{Ma}=0.1$, ${\rm{Re}}_{L}= U_\infty L/ \nu   =10^5$ where $L=100$ is the length of the flat plate. As shown in Fig.~\ref{fig:test3_mesh}, the physical-space computational domain has a length of $150$ in $x \in [ -50,100]$ and a height of $100$ to diminish the influence of the domain boundary, and the leading edge of the plate is placed at $x=0$. In $x$-direction the computational domain is discretized into $150$ cells, where $100$ cells are in the range $x \in [ 0,100]$ along the flat plate. The increasing rates of the cell size from the leading edge of the plate are around $1.08$ upstream and $1.04$ downstream, with the minimum cell width $\Delta x_{\rm{min}}=0.1$ near the leading edge. In $y$-direction, three different resolutions are adopted and the minimum cell heights are $\Delta y_{\rm{min}}=0.1, 0.06, 0.02$, with the same increasing rate around $1.1$ from the surface of the plate to the far field and the discretization numbers are $48$, $54$, $65$ respectively. The setup of the boundary condition is shown in Fig.~\ref{fig:test3_mesh}, where the top and left boundaries of the domain are set as the freestream condition, the right boundary is set as the outlet condition, on the bottom boundary during the range $x \in [ -50,0]$ ahead of the plate the symmetric condition is applied, and it is notable that the no-slip bounce-back boundary condition (implemented by simply mirroring the variables into the ghost cells), but not the diffuse reflection boundary condition discussed in Section~\ref{sec:scheme1bc}, is imposed on the surface of the plate. For the discrete velocity-space domain, the range is $\left\{ {{{({u_x},{u_y})}^T}|{u_x} \in [ - 6{a_\infty},6{a_\infty}],{u_y} \in [ - 6{a_\infty},6{a_\infty}]} \right\}$ where $a_\infty$ is the freestream acoustic velocity, and it is discretized into $20 \times 20$ uniform cells.

\begin{figure}
\centering
\subfigure[Scheme I]{%
\includegraphics[width=0.49\textwidth]{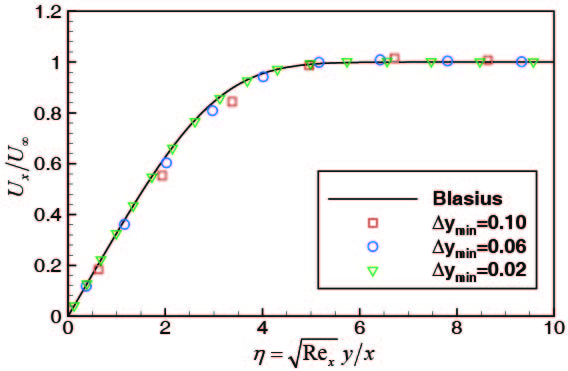}\hspace{0.02\textwidth}%
\includegraphics[width=0.49\textwidth]{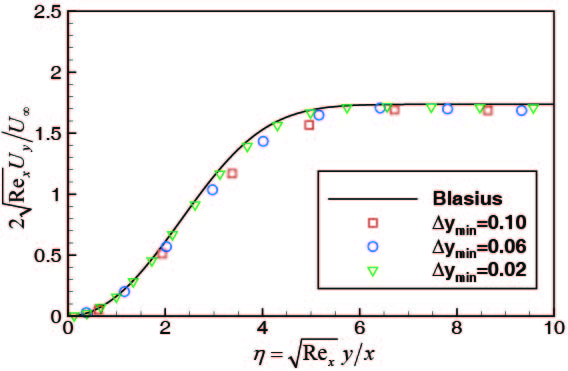}}\\%
\subfigure[Scheme II]{%
\includegraphics[width=0.49\textwidth]{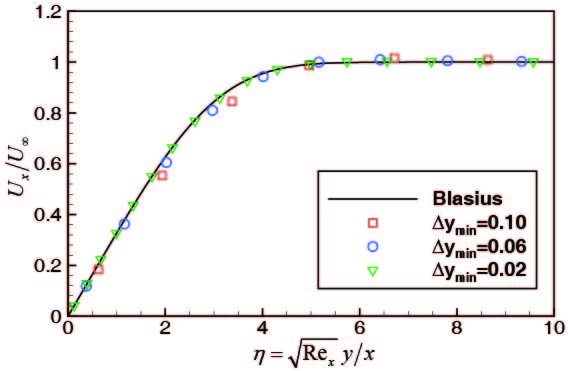}\hspace{0.02\textwidth}%
\includegraphics[width=0.49\textwidth]{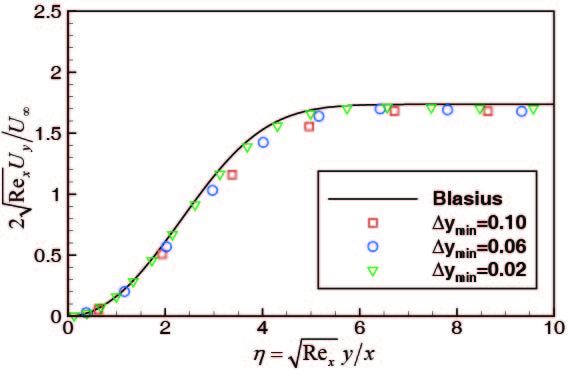}}\\%
\subfigure[GKS]{%
\includegraphics[width=0.49\textwidth]{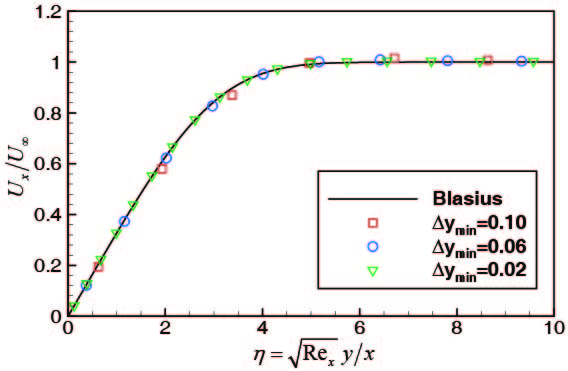}\hspace{0.02\textwidth}%
\includegraphics[width=0.49\textwidth]{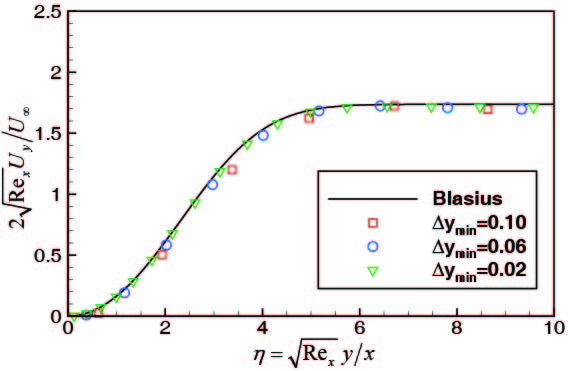}}\\%
\subfigure[DVM\label{fig:test3_cmpvec_dvm}]{%
\includegraphics[width=0.49\textwidth]{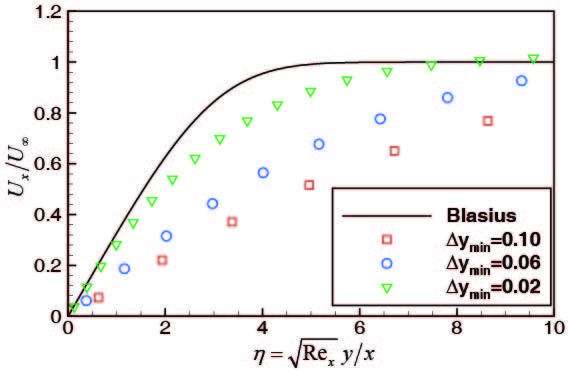}\hspace{0.02\textwidth}%
\includegraphics[width=0.49\textwidth]{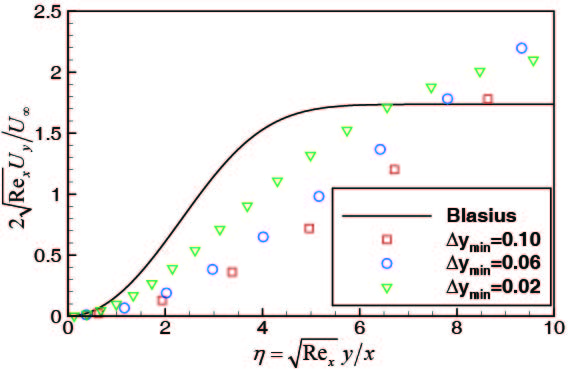}}
\caption{\label{fig:test3_cmpvec}Horizontal (left) and vertical (right) velocity profiles at $x \approx 6.4346$ obtained from different numerical methods under different mesh resolutions.}
\end{figure}

For purpose of comparison, the simulations under the same conditions have also been performed by GKS \cite{xu2001gas}, which is a numerical scheme working in the hydrodynamic (Navier-Stokes) scale, and the conventional DVM (more exactly, the conventional DVM of first-order temporal accuracy discussed in Section~\ref{sec:method}), which is a numerical scheme working in the kinetic (mean-free-path) scale. The results of the velocity profiles at $x \approx 6.4346$ obtained from the four methods are shown in Fig.~\ref{fig:test3_cmpvec}. It is shown that for the scheme I, the scheme II and GKS, they can give roughly accurate velocity profiles even under the coarsest mesh where the minimum mesh height is $\Delta y_{\rm{min}}=0.1$ and there are only $4$ cells in the boundary layer. The deviations of the vertical velocity profiles are relatively larger under such a coarse resolution. Besides, the vertical velocity profiles obtained from these three methods under all resolutions fall below the analytic solution in the far field, which may due to the singularity of the flow at the leading edge because the cross section $x \approx 6.4346$ is close to there. Generally speaking the three methods are in the same level of accuracy under all of the mesh resolutions for this test case. In contrast, for the conventional DVM, it fails to give an acceptable result even under the finest mesh, and all of its results deviate largely from the analytical solution, showing a much thicker boundary layer and suggesting excessive numerical dissipation.

In conclusion, both the multiscale DVM (the scheme II) and the scheme I can give an accurate result in the continuum flow regime just as an NS-equation-based scheme under the same resolution, and the excessive numerical dissipation for the conventional DVM occurring in the low-cell-Kn-number case does not exist in the multiscale DVM (the scheme II) as well as the scheme I.

\clearpage

\section{Conclusions}\label{sec:conclusions}
In this paper, attempt has been made to improve the conventional DVM into a multiscale scheme in the finite volume framework. Instead of reconstructing the discrete distribution function at the cell interface and then calculating the macroscopic flux through integration, the paper presents a DVM-based moment method (the scheme I) where the nonequilibrium flux is firstly obtained from the integration of the discrete distribution function at the cell center and then the flux at the interface is reconstructed, which overcomes the excess numerical dissipation of the conventional DVM in the low-cell-Kn-number case. On this basis the paper modifies the scheme I by a simple treatment and presents the multiscale DVM (the scheme II), which achieves the multiscale property and can work properly in both hydrodynamic scale and kinetic scale just as the multiscale methods UGKS and DUGKS.

Several numerical tests have been performed to investigate the performance of the schemes presented in the paper. In the unsteady test case of Sod's shock tube, the multiscale DVM (the scheme II) shows good consistence with UGKS in all flow regimes while the scheme I only works properly in the continuum regime. In the cavity flow simulation, the multiscale DVM can give results agreeing very well with the results of GKS (in continuum regime) and UGKS, and the scheme I can also give acceptable results except some minor deviations in the rarefied cases. In the test case of laminar boundary layer flow over a flat plate, both the two schemes presented in the paper show precision comparable to the second-order NS-equation-based scheme, and the results are much better than the diffusive results obtained from the conventional DVM.

In conclusion, the multiscale DVM presented in this paper has a good multiscale property and can be applied to gas flow simulations in all flow regimes. More important, the key idea of integrating the discrete distribution function at the cell center, but not the reconstructed distribution function at the cell interface, to calculate the nonequilibrium moments provides another strategy about constructing a multiscale kinetic scheme, and also throws light on the mechanism of the multiscale kinetic scheme.

\clearpage


\bibliographystyle{yuan_mdvm_arxiv}
\bibliography{yuan_mdvm_arxiv}

\begin{thebibliography}{10}

\bibitem{Goldstein1989Investigations}
D.~Goldstein, B.~Sturtevant, and J.~E. Broadwell.
\newblock Investigations of the motion of discrete-velocity gases.
\newblock \emph{Progress in Astronautics and Aeronautics}, 1989.
\newblock 117:100--117.

\bibitem{Yang1995Rarefied}
J.~Y. Yang and J.~C. Huang.
\newblock Rarefied flow computations using nonlinear model {Boltzmann}
  equations.
\newblock \emph{Journal of Computational Physics}, 1995.
\newblock 120(2):323--339.

\bibitem{Mieussens2000Discretev}
L.~Mieussens.
\newblock Discrete-velocity models and numerical schemes for the
  {Boltzmann-BGK} equation in plane and axisymmetric geometries.
\newblock \emph{Journal of Computational Physics}, 2000.
\newblock 162(2):429--466.

\bibitem{li2004Study}
Z.-H. Li and H.-X. Zhang.
\newblock Study on gas kinetic unified algorithm for flows from rarefied
  transition to continuum.
\newblock \emph{Journal of Computational Physics}, 2004.
\newblock 193(2):708--738.

\bibitem{Titarev2007Conservative}
V.~A. Titarev.
\newblock Conservative numerical methods for model kinetic equations.
\newblock \emph{Computers \& Fluids}, 2007.
\newblock 36(9):1446--1459.

\bibitem{Bird1994Molecular}
G.~A. Bird.
\newblock \emph{Molecular gas dynamics and the direct simulation of gas flows}.
\newblock Clarendon Press, 1994.

\bibitem{Xu2010A}
K.~Xu and J.~C. Huang.
\newblock A unified gas-kinetic scheme for continuum and rarefied flows.
\newblock \emph{Journal of Computational Physics}, 2010.
\newblock 229(20):7747--7764.

\bibitem{Xu2015Direct}
K.~Xu.
\newblock \emph{Direct modeling for computational fluid dynamics: construction
  and application of unified gas-kinetic schemes}.
\newblock World Scientifc, 2015.

\bibitem{guo2013discrete}
Z.~Guo, K.~Xu, and R.~Wang.
\newblock Discrete unified gas kinetic scheme for all Knudsen number flows:
  Low-speed isothermal case.
\newblock \emph{Physical Review E}, 2013.
\newblock 88(3):033305.

\bibitem{guo2015discrete}
Z.~Guo, R.~Wang, and K.~Xu.
\newblock Discrete unified gas kinetic scheme for all Knudsen number flows.
  {II}. {Thermal} compressible case.
\newblock \emph{Physical Review E}, 2015.
\newblock 91(3):033313.

\bibitem{chen2016simplification}
S.~Chen, Z.~Guo, and K.~Xu.
\newblock Simplification of the unified gas kinetic scheme.
\newblock \emph{Physical Review E}, 2016.
\newblock 94(2):023313.

\bibitem{yang2018animproved}
L.~Yang, C.~Shu, W.~Yang, Z.~Chen, and H.~Dong.
\newblock An improved discrete velocity method (DVM) for efficient simulation
  of flows in all flow regimes.
\newblock \emph{Physics of Fluids}, 2018.
\newblock 30(6):062005.

\bibitem{yang2018improved}
L.~Yang, Z.~Chen, C.~Shu, W.~Yang, J.~Wu, and L.~Zhang.
\newblock Improved fully implicit discrete-velocity method for efficient
  simulation of flows in all flow regimes.
\newblock \emph{Physical Review E}, 2018.
\newblock 98(6):063313.

\bibitem{yang2019improved}
L.~Yang, C.~Shu, W.~Yang, and J.~Wu.
\newblock An improved three-dimensional implicit discrete velocity method on
  unstructured meshes for all Knudsen number flows.
\newblock \emph{Journal of Computational Physics}, 2019.
\newblock 396:738--760.

\bibitem{su2019can}
W.~Su, L.~Zhu, P.~Wang, Y.~Zhang, and L.~Wu.
\newblock Can we find steady-state solutions to multiscale rarefied gas flows
  within dozens of iterations?
\newblock \emph{Journal of Computational Physics}, 2020.
\newblock 407:109245.

\bibitem{pieraccini2012microscopically}
S.~Pieraccini and G.~Puppo.
\newblock Microscopically implicit--macroscopically explicit schemes for the {BGK} equation.
\newblock \emph{Journal of Computational Physics}, 2012.
\newblock 231(2):299--327.

\bibitem{bhatnagar1954model}
P.~L. Bhatnagar, E.~P. Gross, and M.~Krook.
\newblock A model for collision processes in gases. {I. Small} amplitude
  processes in charged and neutral one-component systems.
\newblock \emph{Physical Review}, 1954.
\newblock 94(3):511.

\bibitem{shakhov1968generalization}
E.~Shakhov.
\newblock Generalization of the {Krook} kinetic relaxation equation.
\newblock \emph{Fluid Dynamics}, 1968.
\newblock 3(5):95--96.

\bibitem{venkatakrishnan1995convergence}
V.~Venkatakrishnan.
\newblock Convergence to steady state solutions of the {Euler} equations on
  unstructured grids with limiters.
\newblock \emph{Journal of Computational Physics}, 1995.
\newblock 118(1):120--130.

\bibitem{chapman1990mathematical}
S.~Chapman, T.~G. Cowling, and D.~Burnett.
\newblock \emph{The mathematical theory of non-uniform gases: an account of the
  kinetic theory of viscosity, thermal conduction and diffusion in gases}.
\newblock Cambridge university press, 1990.

\bibitem{xiong2015high}
T.~Xiong, J.~Jang, F.~Li, and J.~M.~Qiu.
\newblock High order asymptotic preserving nodal discontinuous {Galerkin} {IMEX} schemes for the {BGK} equation.
\newblock \emph{Journal of Computational Physics}, 2015.
\newblock 284:70--94.

\bibitem{jang2015high}
J.~Jang, F.~Li, J.~M.~Qiu, and T.~Xiong.
\newblock High order asymptotic preserving {DG-IMEX} schemes for discrete-velocity kinetic equations in a diffusive scaling.
\newblock \emph{Journal of Computational Physics}, 2015.
\newblock 281:199--224.

\bibitem{xiong2017hierarchical}
T.~Xiong and J.~M.~Qiu.
\newblock A hierarchical uniformly high order {DG-IMEX} scheme for the {1D} {BGK} equation.
\newblock \emph{Journal of Computational Physics}, 2017.
\newblock 336:164--191.

\bibitem{xu2001gas}
K.~Xu.
\newblock A gas-kinetic {BGK} scheme for the {Navier-Stokes} equations and its
  connection with artificial dissipation and {Godunov} method.
\newblock \emph{Journal of Computational Physics}, 2001.
\newblock 171(1):289--335.

\bibitem{mandal1994kinetic}
J.~Mandal and S.~Deshpande.
\newblock Kinetic flux vector splitting for {Euler} equations.
\newblock \emph{Computers \& fluids}, 1994.
\newblock 23(2):447--478.

\bibitem{yuan2019conservative}
R.~Yuan and C.~Zhong.
\newblock A conservative implicit scheme for steady state solutions of diatomic
  gas flow in all flow regimes.
\newblock \emph{Computer Physics Communications}, 2019.
\newblock page 106972.

\bibitem{yuan2019multi}
R.~Yuan and C.~Zhong.
\newblock A multi-prediction implicit scheme for steady state solutions of gas
  flow in all flow regimes.
\newblock \emph{arXiv preprint arXiv:1905.06629}, 2019.

\bibitem{Li2005Application}
Q.~Li, S.~Fu, and K.~Xu.
\newblock Application of gas-kinetic scheme with kinetic boundary conditions in
  hypersonic flow.
\newblock \emph{AIAA Journal}, 2005.
\newblock 43(10):2170--2176.

\bibitem{Zhu2016Implicit}
Y.~Zhu, C.~Zhong, and K.~Xu.
\newblock Implicit unified gas-kinetic scheme for steady state solutions in all
  flow regimes.
\newblock \emph{Journal of Computational Physics}, 2016.
\newblock 315:16--38.

\end{thebibliography}

\clearpage

\clearpage
\renewcommand{\multirowsetup}{\centering}

\end{document}